\documentclass[14pt]{extarticle}
\usepackage{amssymb,amsmath,amsfonts,amsthm,graphicx,psfrag}
\usepackage{cmap}
\usepackage{graphicx}
\usepackage[left=2.25cm,right=2.25cm,top=2.25cm,bottom=2.25cm]{geometry}

\title{   Scalar mesons in the  chiral theory with quark degrees of freedom}
\author{ M.S. Lukashov and  Yu.A. Simonov \\
 Institute  for Theoretical and Experimental Physics,\\ NRC ``Kurchatov Institute''
 \\
Moscow, 117218 Russia}
\date{\today\thanks{v5 for arXiv (accepted to PRD)}} 

\newcommand{\beq}{\begin{eqnarray}}
 \newcommand{\eeq}{\end{eqnarray}}
\newcommand{\be}{\begin{equation}}
 \newcommand{\ee}{\end{equation}}

\def\fun#1#2{\lower3.6pt\vbox{\baselineskip0pt\lineskip.9pt
\ialign{$\mathsurround=0pt#1\hfil ##\hfil$\crcr#2\crcr\sim\crcr}}}

\newcommand{{\SD}}{\rm SD}

\newcommand{{\Mc}}{\mathcal{M}}

\newcommand{\vex}{\mbox{\boldmath${\rm x}$}}
\newcommand{\vey}{\mbox{\boldmath${\rm y}$}}
\newcommand{\ver}{\mbox{\boldmath${\rm r}$}}

\newcommand{\veP}{\mbox{\boldmath${\rm P}$}}
\newcommand{\vep}{\mbox{\boldmath${\rm p}$}}

\newcommand{\lan}{\langle}
\newcommand{\ran}{\rangle}

\begin{document}
\maketitle
\begin{abstract}
 The Chiral Confining Lagrangian, based on the  chiral theory with  quark degrees of freedom, is used to study the spectroscopy of scalar mesons. The formalism does not  contain  arbitrary fitting parameters and takes into account  infinite number of  transitions from meson-meson to quark-antiquark states. Starting from known $q\bar q$ poles  the transition  coefficients ensure the strong shift of the poles for the  $\pi\pi$ and much smaller shift for the $K\bar K$ systems. The resulting amplitudes $f_{\pi\pi}$ and $f_{K\bar K}$ are calculated in terms of the $q\bar q$ and the free meson   Green's functions. With the account of the $\pi\pi/K\bar{K}$ channel coupling one obtains two resonances: a wide resonance $E_1$ in the range  500-700 MeV and narrow  $E_2 $ near 1 GeV,  which can be associated with  $f_0(500)$ and $f_0(980)$. A similar analysis, applied to the $I=1$ channel, shows that in this case two very close poles  in different  sheets appear near $E=980$ MeV, which  can be associated with  the $a_0 (980)$ resonance.  The obtained $\pi\pi$ interaction amplitudes, $\operatorname{Re}  f_{\pi\pi} (E)$ and $\operatorname{Im} f_{\pi\pi}(E)$ are compared with the known data.
 \end{abstract}

 \section{Introduction}
 Scalar mesons are in the  center of experimental and theoretical interests for  a long time (see summary of experimental data in Ref. \cite{1} and a large   amount of  information about the scalars in the reviews \cite{2,3,4,5,6,6a}, and recent  comprehensive analysis in \cite{7*,8*}). The theoretical explanation of the  scalar spectrum has faced difficulties and required the  development of different approaches, like the tetraquark model \cite{7}, the molecular approach \cite{9}, and the QCD  sum rules \cite{10}, as
 well as lattice calculations \cite{11} (see recent study in \cite{12'}).


It is clear, that in QCD any meson state can be represented as a series $M = c_1\,(q\bar{q}) + c_2\,(q\bar{q})^2 + ...$, where
 higher terms can be transformed into mesons as $(q\bar{q})^n=m_1\,m_2\,... $
 Mesons with nonzero $q\bar q$ component can be called standard, while those with $c_1=0$ --nonstandard or exotic.
 For standard mesons the original $q\bar{q}$ pole can be shifted due to $q\bar{q}$-$mm$-$q\bar{q}$ interaction, as it is known from comparison with experiment. However, the $m_1m_2$ interaction can be strong enough to produce bound states and resonances as it happens in nuclear physics.
 In what follows we shall study scalar resonances in QCD, starting from the standard $q\bar{q}$ component and describing the physical scalar resonance as the result of multiple $q\bar{q}$-$m_1m_2$ transitions. In principle this approach
 is not new and has been worked out in \cite{A,B,C,D,13',13",13'",20*}, where transitions have been properly parametrized.  On another hand, one can have additional poles due to $m_1m_2$ interaction. The latter can be introduced in the framework of the meson-meson interaction in the unitarized chiral perturbation theory (also on top of quark model resonances) \cite{E,F,G,H,I,J,K}, and here e.g. $f_0(500)$ becomes heavier and more narrow, when $mm$ interaction is suppressed \cite{F,G}. For the results of the unitarized chiral perturbation theory and inclusion of NLO terms see \cite{I,J} and the review paper \cite{7*}.
 With all that one can stress that the Chiral Perturbation Theory is not necessary for obtaining these results and one can use the dispersive methods and the data to obtain a good explanation of $f_0(500)$ and other resonances, see e.g. \cite{19**, 39, 40}.

As it is, the situation with the scalar mesons, and first of all, with lowest scalar mesons, is still unclear and calls for new ideas. As one can see in \cite{1}, Table 2, the conventional opinion considers the resonances $a_0 (1450)$ and $f_0(1370)$ as the lowest ${^3}P_0$ states for $I=1,0$ respectively. On the other hand, numerous   calculations of  the lowest $^3P_0$  $q\bar q$ states with realistic $q\bar q$ interaction, including spin-dependent forces refer  to $a_0(980) (f_0 (980))$ as the lowest $^3P_0$ states, see e.g.  \cite{29}, while $a_0(1450)$ might be only  connected to the first excited state.

There is no general consensus on  the lowest states ($f_0(500)$, $f_0(980), a_0(980)$) in the modern approaches, including the  attempts to derive these states in the  molecular  or tetraquark approaches. Unfortunately also in this latter  approach    a recent lattice calculation \cite{L} of the  $a_0(980)$ state with account of tetraquark ($q^2\bar q^2)$ contribution does not show any explicit influence of the latter on the lowest states, thus calling for a new dynamics   as a possible source of $f_0(500), f_0(980), a_0(980)$.

It  is the purpose of the present paper to suggest a new approach to the solution of this problem and to demonstrate a new quark-chiral dynamics, which  might explain  the origin of the  lowest scalar states. The essence of the method is as follows.

The main problem of the  most part of approaches to the scalar mesons  from our point of view is  the imbalance in the treatment of quark and meson d.o.f. In  reality meson-meson ($\varphi\varphi$, e.g. $\pi\pi, K\bar K$) and $q\bar q$ d.o.f. have to be considered on equal footing, since both  can transform into each other at any
moment of time. Moreover, the $q\bar q$ poles are accurately  predicted at the proper places by  the relativistic QCD theory with  scalar confinement and gluon exchanges  \cite{13,21*,27*}  in all  channels \cite{27**,14} and in  many cases they are observed in  experiment shifted by 50-80 MeV or less.

Therefore, the $q\bar q$ poles should be seen in experiment as the $\varphi\varphi$ resonances, shifted or not shifted. On another hand one may think of some (or all) $ \varphi \varphi $ resonances as produced by the $\varphi\varphi$ interaction, e.g. by the unitarized chiral dynamics , where $q\bar q$ dynamics does not play any role. Instead, we consider the coupled $\varphi\varphi-q\bar q$ system with  the proper $q\bar q$ dynamics and the transition dynamics of $q\bar q$ into $\varphi\varphi$ system, first neglecting the $\varphi\varphi$ interaction and introducing it at the next stage.

Therefore, one needs the formalism of the two-channel $q\bar q$, $\varphi\varphi$ Green's functions, which takes into account any number of  $q\bar q$ -- $\varphi\varphi$, and $\varphi\varphi-q\bar q$  transitions.
In the case of scalar mesons this type of formalism was already exploited in \cite{A,B,C,D,13',13",13'",20*}.In the case of the heavy quarks this formalism, considering $Q\bar Q$ and $(Q\bar q)+(\bar Qq)$ channels nonrelativistically, was suggested in \cite{28'}, and was called the Cornell formalism.  It was used in \cite{28''} to discover the nature of the resonance $X(3872),$ with $c\bar c$ $2^3P_1$ state transforming into $D D^*$, via string breaking mechanism, which finally brings it to the $D_0D^*_0$ threshold at 3872 MeV \cite{28'''}.

We shall generalize the Cornell formalism, making it relativistic and multichannel, when one $q\bar q$ state can transform into several $\varphi\varphi $ states, and we shall neglect at the first stage the interaction between white $\varphi\varphi$
 mesons.

The full analysis of the scalars requires the multichannel approach to the problem, where
several quark-antiquark ($q\bar q$) channels are present
together with two or more Nambu-Goldstone boson channels ($(\varphi\varphi)$ channels). Therefore, 
complete formulation  requires the knowledge of 1) the Green's
functions both in $q \bar q$ and $\varphi\varphi$ channels; 2) the transition matrix
elements between the channels. Without explicit knowledge of these entries one faces the multi-parameter and multi-channel
situation with hardly possible informative output.

The treatment of  the first point -- the spectral representation of the $q\bar q$
Green's function with  accurate calculation of one-channel $q\bar q$ poles
and couplings, can be  done in the framework of  the Field Correlator
Method (FCM) (see \cite{21*,27*} for  reviews  and \cite{27**} for recent calculations
in  different channels). The $\varphi\varphi$ Green's function in the initial one-channel set-up will be
studied here, assuming that it can be
replaced  by  the free two-body propagators and possible resonances  exist only due to channel coupling,
in particular, with the $q\bar q$ channels, and here the problem 2) becomes a basic point in  a new approach.

Indeed, in the heavy quarkonia the channel coupling with the heavy mesons is  described by the string breaking mechanism (sometimes with emission of pions), which brings  in the resonance shift of $O(50-100)$ MeV). In the case of scalar mesons the $\varphi\varphi$ channel contains  chiral mesons and the transition process from $\varphi\varphi$ to $q\bar q$ and back  requires a different approach.

During  the last 15 years  one of the authors has succeeded to derive the Chiral Confining Lagrangian
(CCL) - the  powerful tool for the  study of chiral effects  in connection with quark d.o.f. \cite{23, 23*}.
The  latter is actually an extension of the standard
 Chiral Lagrangian, which contains both the quark and chiral d.o.f. and  tends  to the standard  Chiral Lagrangian \cite{16} when quark d.o.f. are neglected; all coefficients of CCL are easily
calculated, as it was done in \cite{23*} in the  order of $p^4$. Moreover, the basic factors, like $f_\pi, f_K$, are
calculated within this method \cite{25}. The only basic parameter, $M{(\lambda)}=\sigma \lambda$, which appears due
to confinement, is  a fixed quantity, defined by the  transition radius $\lambda$. The latter is calculated at the stationary point and is expressed via string tension $\sigma$ and masses \cite{X} and as a result our method does not contain any fitting parameters. In this way the  CCL method allows to find analytically all entries 1) and 2), while the scalar decay constants $f_s$ are calculated  in the same way as $f_{\pi},\,f_K$ within the FCM, using the spectral representation of the Green's function.

In principle, our method gives a possibility of treating any process with
multiple $q\bar q$ and any number of $\varphi\varphi$ channels; the advantage of
using  the CCL is that for scalar mesons all transition coefficients are known.
In the case of a single $\varphi\varphi$ and  a single $q\bar q$ channel our results can be written in the form, comprising
the  Breit-Wigner resonance, similarly to results in Refs.~\cite{A,B,C,D,13',13",13'",20*}. However, in the case of multiple
$\varphi\varphi$ channels more  complicated expressions are obtained,  using the $K$-matrix approach.

As will be seen, the essence of our approach is the  summation of the infinite re-scattering series with multiple transitions between $\varphi\varphi$  and $q\bar  q$ states, which yields several poles. In this  way we obtain two poles  in the regions of  $f_0$ (500) and $f_0(980)$, which finally obtain realistic positions when $\varphi\varphi$ interaction is taken into account.

The paper is organized as follows. In the next section the   general structure of the coupled-channel Green's function for a scalar meson is derived from CCL, and we define basic quantities 1), 2)  in terms of known standard coefficients. In
section 3 we discuss the $q\bar q$ Green's functions in the spectral form and the free $\varphi\varphi$ Green's functions and use
the decay constants and the pole masses from  the known confining,   gluon exchange and
spin-dependent interaction. Note, that this calculation does not use any
parameters, beyond the string tension, the current quark masses, $\Lambda_{QCD}$,  and $M(\lambda)$.
In section 4 we discuss the resulting  $\varphi\varphi|q\bar q$ Green's function and find the  physical $\varphi \varphi$
amplitudes ($ \pi \pi$ and $K\bar K$),   containing two resonances, which can be  associated with $f_0(500)$ and $f_0(980)$
In section 5  these  results are augmented by the calculation of the real and imaginary parts of the $\pi\pi $ amplitude, which are  qualitatively similar to the  results, obtained  from theory  and experiment at least for $E>500$ MeV. We demonstrate in Fig. 5 and 6 that by a proper modification of the $\pi\pi$ Green's function one is able to reproduce these data, and the resulting $\pi\pi$ resonance is made closer to the experiment. We also show
that in the case  of the isospin $I=1$ our method gives a different picture of two nearby poles within 50 MeV in different sheets  for $a_0(980)$. Section 6 contains discussion and  an outlook.

\section{Coupled channel equations for the scalars from the  Chiral Confining Lagrangian }

In what  follows we are  using the   Chiral Confining Lagrangian
(CCL) \cite{23, 23*} with the scalar external currents $s_0 (x)$ and
$s_a(x)\lambda^a\equiv \hat s$ for isospin $I=0$ and $I=1$, respectively.

\be L_{CCL} =-N_c \operatorname{tr} \log (\hat \partial+ \hat m + s_0 + \hat s + M\hat
U),\label{1}
\ee
In Eq. (\ref{1}) $\hat\partial$ implies $\dfrac{\partial }{\partial x_\mu}\,\gamma_\mu$ and $\hat{U}$ is the  standard  chiral operator,
\be
\hat U = \exp (i \gamma_5 \hat\varphi),~~ \hat \varphi = \frac{\varphi_a \lambda_a}{f_a}, \label{2}\ee
\be
\hat \varphi = \sqrt{2}\left(\begin{array}{ccc} \frac{1}{f_\pi} \left( \frac{\eta}{\sqrt{6}} +
\frac{\pi^0}{\sqrt{2}}\right), &\frac{\pi^+}{f_\pi},&
\frac{K^+}{f_K}\\
\frac{\pi^-}{f_\pi} & \left( \frac{\eta}{\sqrt{6}}-
\frac{\pi^0}{\sqrt{2}}\right) \frac{1}{f_\pi},& \frac{K^0}{f_{K^0}}\\
\frac{K^-}{f_K},& \frac{\bar K^0}{f_{K^0}}, & - \frac{2\eta}{\sqrt{6}
f_\pi}\end{array}\right)\label{3}\ee

In (\ref{2}) $\lambda_a$ are the Gell-Mann matrices, $\operatorname{tr}\lambda_a \lambda_b=
2\delta_{ab}$.
One can consider CCL in (\ref{1}) as a generating functional for different vertices and Green's functions.
Indeed, omitting $s_0$ and $s$, one can expand, as in \cite{23,23*} in powers of $\hat{U}^+\Lambda(\hat{\partial}+\hat{m})(\hat{U}-1)=\eta$ which dimensionless and yields an expansion in $S\hat{\partial}\hat{\phi}$, which is an expansion in quark loops (here $S=i\Lambda$ is the quark propagator) times derivative of chiral field $\phi$ -- this gives $O(p^4)$ terms in good agreement with standard calculations, and the same expansion yields standard GMOR relation -- see \cite{23*} for details. In our case we need another expansion in powers of dimensionless quantity, $\Lambda M (U-1)$, which is done as follows.

Using the scalar currents $s_0$, $\hat{s}$, one can generate the scalar
Green's functions $G^s_{q\bar q}, G^s_{\varphi\varphi}$.
$$ L_{CCL} =-N_c \operatorname{tr} \log (\Lambda^{-1}  + s_0 +\hat s+M(\hat U-1)) =- N_c \operatorname{tr} \log \Lambda^{-1}
(1+\Lambda (s_0 +\hat s + M(\hat U-1)))=$$ \be =\frac{N_c}{2} \operatorname{tr} \left\{(\Lambda
(s_0 + \hat s) \Lambda  (s_0+\hat s))+...\right\} = \frac{N_c}{2} (
{G^{s_0}_{q\bar q}} +  {G^{\hat s}_{q\bar q}} ) +...\label{4}\ee
Here $\Lambda = \frac{1}{\hat \partial + \hat m + M }$. The corresponding  diagram is shown in Fig.1
One can  write $G^s_q (x,y) = \operatorname{tr}\left (\bar s (x)  g_{q\bar q} (x,y) s(y)\right)$ , where  $g_{q\bar q}$ will be used later.

\begin{figure}[!htb]
\begin{center}
\includegraphics[angle=0,width=5 cm]{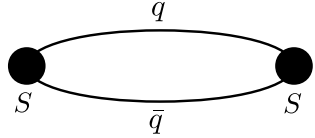}
\caption{ The scalar $q\bar q$ Green's function $G_{q\bar q}$.} 
 \end{center}

\end{figure}

On the other hand,  expanding the CCL (\ref{4}) in  powers of $\Lambda M (\hat
U -1) \equiv \xi$, one obtains another term in the second order in $\xi$,
\be
\Delta L = - N_c \operatorname{tr} \Lambda  s \Lambda   M\frac{\hat \varphi^2}{2}, ~~ s = s_0
+\hat s;\label{5}\ee
which corresponds to the diagram of Fig. 2.
Note, that in this way we can obtain the vertices for all chiral decays of any $q\bar q$ state, e.g. $a_2$ meson decaying into $3\pi$ etc. 

\begin{figure}[!htb]
\begin{center}
\includegraphics[angle=0,width=7 cm]{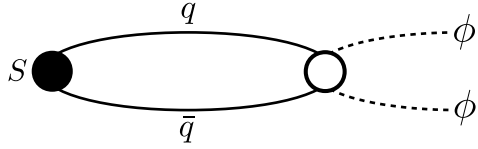}
\caption{ The scalar $q\bar q$ Green's function with the emission of the chiral mesons.} 
\end{center}
\end{figure}

In (\ref{1})  the confining kernel $M(r)$ enters either inside the propagating $q\bar q$ system, in which case it is equal to the confining potential, $M(r) = \sigma r$, or else it appears at the vertex of the $q\bar q$ Green's function, connecting it to the $ \varphi\varphi$ Green's function. In this case the vertex $M(r)$ is taken  at the effective distance $\lambda, ~ M=M(\lambda) =\sigma \lambda$. One can consider this distance $\lambda$ as the spatial width of the transition vertex, connecting $\varphi\varphi$ and $q\bar q$ channels,  see Fig.3. In the case of one chiral meson    we take it  approximately equal to the correlation length in the confining vacuum, $\lambda \approx 0.2$ fm, yielding $M(\lambda)=0.15$ GeV. As a check of this approximation, this value was used to calculate $f_\pi$ and $f_K$ \cite{25} in good agreement with  experimental and lattice data and therefore we  shall consider   $ \lambda$ in the range  $(0.2\div 0.3) $  fm (or  $1\div 1.5$ GeV$^{-1}$) in what follows. This factor $M(\lambda)$ appears to be the only parameter of our quark-chiral approach in (\ref{1}) in addition to the quark masses $m_q$ and string tension $\sigma$.

 From (\ref{5}) one can find the basic quantity, which will be used below, -- the transition element $V_{q\bar q  \varphi\varphi}$ which joins the $q\bar q$ Green's function $g_{q\bar q}$ and the $\varphi\varphi$ Green's function $g_{\varphi\varphi}$, see Fig.4 and its definition below. At this point it is important to understand which kind of the $q\bar q$ Green's function is needed to join it with the $g_{\varphi\varphi}$, i.e. to annihilate at one vertex $q\bar q$ and create at this vertex two mesons $\varphi\varphi$. One  clearly needs $g_{q\bar q} (x,y) \sim (S_q(x,y) S_{\bar q} (x,y)$, where $S_q(x,y)$ is the quark Green's function, but with the definite total momentum, i.e. $g_{q\bar q} (P) = \int d^4 (x-y) e^{iP(x-y)} \operatorname{tr} (S_q (x,y) S_{\bar q}(x,y))$; originally $g_{q\bar q} (x,y)  $ should be connected with $g_{ \varphi\varphi}$ at the same point $x$ or $y$ and finally with $g_{\varphi\varphi}(P)$. However, $g_{\varphi\varphi}(P)$ is divergent in its real part, which implies that the transition from $q\bar q$  to $\varphi\varphi$ occurs not in one  point, but at some distance between $q$ and $\bar q$, namely, at the same distance between $\varphi$ and $\varphi$ which we call $r_0 \sim \lambda \sim 0.2 $ fm -- the transition radius, which is shown in Fig.3.

\begin{figure}[!htb]
\begin{center}
\includegraphics[angle=0,width=5 cm]{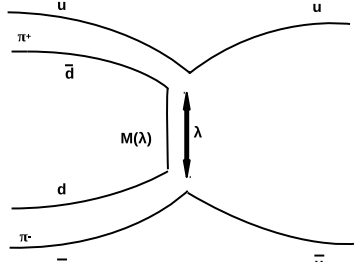}
\caption{The transition region ($q\bar q|\varphi\varphi)$ with the spatial distance $\lambda$ between the  constituents.} 
 \end{center}

\end{figure}
 It is important  that at this moment the  $M(r)$ becomes $M(\lambda)=\sigma \lambda\approx 0.15$  GeV, and $ \operatorname{Re} g_{\varphi\varphi}$ should have an initial and final  $\varphi-\varphi$  distance $\lambda$.  As will be shown below, this  transition radius does not change much the $g_{q\bar q} (\lambda)$, which is anyhow convergent at $\lambda=0$, but the variation of $\operatorname{Re} g_{\varphi\varphi}(\lambda)$ can be taken into account. In this approximation the total scalar Green's function can be written as

 \be G^s = {g^{s}_{q\bar q}}+ {g^{~s} _{q\bar
q}}V  g^s_{\varphi\varphi}V{g^{s} _{q\bar q}} +...= {g^{s} _{q\bar q}}
\frac{1}{1-V g_{\varphi\varphi}^sV{g^{s} _{q\bar q}}}\label{6}\ee
Here $V \equiv  V_{q\bar q/ \varphi \varphi}$ can be  found  from (\ref{5}), see below.

As it is  seen from  (\ref{5}), the transition coefficient $V$ is proportional to  $\frac{M(\lambda)}{f^2_\varphi}, ~~ \varphi=\pi,K,$ and also to the quark decay constant of the scalar meson $f_s^{(n)}, ~~( n=1,2...)$ to be  found  below.

Finally, to define how $V$ depends on isotopic indices, one can according to (\ref{5}), project $\frac{\hat \varphi^2}{2} $ on a given isotopic state with $I=0$ or 1.

 \be \operatorname{tr} \left(s_0
 \frac{\hat \varphi^2}{2} \right) = s_0 (a_{11} + a_{22} +a_{33});\label{20}\ee

$$ \operatorname{tr} \left(s_i \lambda_i\frac{\hat \varphi^2}{2}\right) =  a_{11} \left(s_3
+\frac{1}{\sqrt{3}} s_8\right) +a_{22} \left(-s_3 +\frac{1}{\sqrt{3}}
s_8\right) +a_{12} \left(s_1+i s_2\right)+$$\be +a_{21} \left(s_1-i s_2\right)-
a_{33} \cdot \frac{2}{\sqrt{3}} s_8\label{21}\ee
where $a_{ik}$ are

\be  a_{11 }= \frac{1}{f^2_\pi}\left[ \left( \frac{\eta}{\sqrt{6}}+
\frac{\pi^0}{\sqrt{2}}\right)^2 + \pi^+ \pi^- \right] + \frac{K^+
K^-}{f^2_K},\label{22}\ee

\be a_{12} = \frac{2 \eta \pi^+}{f^2_\pi \sqrt{6}} +\frac{K^+ \bar K^0}{f^2_K},
~~a_{21} = \frac{2 \eta \pi^-}{f^2_\pi \sqrt{6}} +\frac{K^0
K^-}{f^2_K},\label{23}\ee

\be a_{22} =  \frac{1}{f^2_\pi} \left[ \left(
\frac{\eta}{\sqrt{6}}-\frac{\pi^0}{\sqrt{2}}\right)^2 + \pi^+\pi^-\right]+
\frac{K^0 \bar K^0}{f^2_K}, \label{24}\ee

\be a_{33} = \frac{K^+ K^-+ K^0\bar
K^0}{f^2_K}+ \frac23 \frac{\eta^2}{f^2_\pi}.\label{25}\ee


Later we shall neglect the   isotopic (SU(3)) dependence of  the
propagators $\Lambda$, apparent  in  the  mass matrices $\hat m$, and take it
in account at the end, since  one can write $ g_{q\bar q} \equiv  g_1= \left( \begin{array} {ll}g_1(n\bar n)&0\\0&
g_1(s\bar s) \end{array}\right).$

\section{The $q\bar q$ Green's functions and the eigenvalues}

To calculate the $q\bar q$ Green's functions we shall use the exact relativistic  formalism, based on the FCM \cite{13} and  essentially exploiting relativistic path integral methods \cite{27*,27**,14,26,27};  at the end we shall compare our results with those obtained in other methods.

The $q\bar q$ Green's function $g^\Gamma_{q\bar q} (x,y) \equiv g_1(x,y)$ with  the vertex $\Gamma$, defining the spin-parity,  can be written as

\be g_1(x,y) = \operatorname{tr} \left( \frac{4Y}{(m_1^2-\hat D^2_1)(m^2_2 - \hat D^2_2)}\right)\label{14}\ee
where
\be 4  Y = \operatorname{tr} [\Gamma(m_1-\hat D_1) \Gamma(m_2 -\hat D_2)].\label{15}\ee
Then using   the relativistic path integral formalism (see \cite{26,27,28} for a review) it can be written in the c.m. system and in the Euclidean time $T$
\be
\int d^3 (\vex-\vey) g_1 (x,y) = \frac{T}{2\pi} \int^\infty_0\frac{d\omega_1}{\omega_1^{3/2}} \int^\infty_0 \frac{d\omega_2}{\omega_2^{3/2}}\lan Y\ran \lan 0|e^{-H(\omega_1, \omega_2,\vep)T}|0\ran.\label{3.3}\ee
Here the c.m. Hamiltonian $H(\omega_1, \omega_2, \vep)$ depends on the virtual energies $\omega_1,\omega_2$ and includes all instantaneous interactions, including spin and angular momentum dependent,

\be H(\omega_1, \omega_2,\vep) = \sum_{i=1,2} \frac{\vep^2+\omega^2_i +m^2_i}{2\omega_i} + V_0(r) +V_{so} + V_T.\label{3.4}\ee

Here $V_0(r) =\sigma r - \frac43 \frac{\alpha_V(r)}{r},~~ V_{so}$ is  the spin-orbit interaction and $V_T$ is the tensor interaction, both in the relativistic form. Neglecting spin terms, one can rewrite the last term in (\ref{3.3}) as
\be \lan 0|e^{H(\omega_1, \omega_2 \vep)T}|0\ran = \sum_{n=0} \varphi^2_n (0) e^{-M_n (\omega_1, \omega_2)T},\label{3.5}\ee
where $\varphi_n (\ver)$ is the wave function. On the other hand one has a general relation

$$ \int g_1 (x,y) d^3(\vex-\vey) = \sum_n \int d^3 (\vex-\vey) \lan 0|j_\Gamma |n\ran \lan n|j_\Gamma|0\ran\times $$
\be  e^{i\veP (\vex-\vey) - M_n T}\ \frac{d^3\veP}{2M_n (2\pi)^3} = \sum_n  \varepsilon_\Gamma \otimes \varepsilon _\Gamma \frac{(M_n f_\Gamma^{(n)})^2}{2M_n} e^{-M_nT}\label{3.6}\ee
This relation allows to calculate the scalar decay constant $f_s^{(n)}$, which is done in Appendix 1.

Note, that using CCL, Eq.(\ref{1}), one would have in (\ref{15}) $m_i+M(\lambda)$ instead of $m_i$, which allows obtaining in the PS case $(\Gamma= \gamma_5)$ the correct decay constants $f_\pi, f_K$, \cite{25}, which otherwise would be zero in the zero quark mass limit.

In (\ref{3.3}) it is convenient to integrate over $d\omega_1, d\omega_2$, using the stationary-point method, and for vanishing quark masses $m_i=0$ one obtains the so-called spinless Salpeter equation; if spin-dependent  interactions are neglected. In the first approximation one has
\be (2\sqrt{\vep^2} + V_0 (r)) \varphi_{nl} (r) = M_{cog} (nl) \varphi_{nl} (r),\label{3.7}\ee
where $M_{cog}$ means the center-of-gravity mass. Later we use only the fundamental parameters: $\sigma =0.182(2)$~GeV$^2$ and $\Lambda_V (n_f=3) =0.465(15)$ GeV, which   are well established  (see \cite{35*} for the definition of $\Lambda_V$ and an accurate perturbative threatment of scalar mesons), and obtain
\be M_{cog} (1P) = 1259 (10)~ {\rm MeV}, ~~ \omega_0 (1P) = 499~{\rm MeV}.\label{3.8}\ee

Doing calculations in the  same  way as in  \cite{27*,27**,14},
 we   give here the resulting mass of the $1\,^3P_0$ state with  account of the tensor and spin-orbit forces
\be M (1^3P_0) = (1259(10)- 214)~{\rm MeV} =1045(10)~{\rm MeV},\label{3.9}\ee
which defines the $q\bar q$ initial mass of $f_0$ and $a_0$, taken below,  as $M_1=1 $ GeV. This mass can be compared with that  obtained by  other groups, where in \cite{30}  $ M(0^{++})=1090$ MeV,  while in \cite{31} $M (0^{++}) =1176$ MeV, and in \cite{32},   $M(0^{++})=970$ MeV.

Note, that the first excited state in the $0^{++}, I=0$  channel is obtained to be $M_2=1474$~MeV \cite{28} and this state can be associated with the $a_0(1450)$.

Finally, we can use (\ref{3.6}) to calculate the full Fourier transform of $g_1(x,y)$ in the Minkowskian time, which yields

 \be \tilde g_1 (P) = \tilde g_1 (E, \veP=0) = \sum_n \frac{(f^{(n)}_s )^2 M^2_n}{M^2_n - E^2}.\label{3.10}\ee

This form with the lowest $n=1 $ will be used below to analyse  the scalars $f_0$; it will be shown that the level $M_1=1045$ MeV generates both $f_0(500)$ and $f_0(980)$ resonances, connected respectively with the $\pi\pi$ and $K\bar K$ Green's functions.

\begin{figure}
\begin{center}
  \includegraphics[width=13cm, ]
  {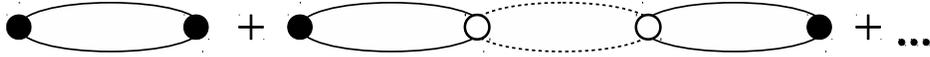}

  \caption{The $\pi\pi$ interaction amplitude in terms of the $q\bar q$ (solid lines) and $\pi\pi$ Green's functions (broken lines). The  filled and empty
circles  denote the transition matrix elements $V_{\pi 1} = V_{1 \pi}$. }
 \end{center}

\end{figure}

We turn to  the structure of the meson-meson Green's function, which at first we take as free two body relativistic Green's function of two scalar particles with the total momentum $\veP=0$ and the total c.m. energy $E$.See Appendix 2 for the detailed discussion. 


Then in the $\varphi\varphi$ channel the free Green's function of $\varphi\varphi$ displaced by a spatial distance
$\lambda$ and averaged over its direction ($S$-wave) brings in an additional 
factor $f_2(|\vep|\lambda)$:

\be g_2 (E) =  \int \frac{f_2(|\vep|\lambda)d^4p}{(2\pi)^4 (p^2-m^2_1)
((P-p)^2-m^2_2) },\label{31}\ee
with $\veP =0; P_0 =E$. Its imaginary part is

\be \operatorname{Im} g_2 (E) =\frac{ \sqrt{(E^2-(m_1+m_2)^2) (E^2-(m_1-m_2)^2)}}
 {16 \pi E^2},\label{32}\ee
One can compare (\ref{32}) with the cut-off integral, where for equal masses $m_1=m_2=m$ one has for  the real part with the cut-off function $f_1(|\vep|\lambda)= \theta (1- |\vep|\lambda), ~~ N=1/\lambda$
\be  \operatorname{Re} g_2(E=2m) = \frac{1}{8\pi^2}\ln \left( \frac{N+ \sqrt{N^2+
m^2}}{m}\right).\label{32a}\ee

Note, that $f_2(|\vep|\lambda)$ is not a cut-off introduced by hand, as $f_1(|\vep|\lambda)$ which will be used below for comparison. Moreover, $f_2$ is a part of a physical amplitude, not violating unitarity in the case of spatial distance, and therefore not producing additional singularities. Indeed, one can see this in the explicit form, since $f_2$ is expanded in even powers of $\vep$.

In the case of the  spatial cut-off $f_2(|\vep|\lambda)$, when the initial and final distances between $\varphi\varphi$ are equal to $\lambda$, we have $f_2(x)=\left( \frac{\sin x  }{x}\right)^2$, and the resulting difference between the two real parts with $f_1$ and $f_2$ is less than 10\% for $\lambda =( 0.5 \div 2)$ GeV$^{-1}$.

Note,  that the spatial cut-off does not introduce branch points into $g_2(E)$ and therefore does not spoil unitarity.

 \section{Analytic structure of physical amplitudes}
We start with the transition coefficient, which we denote $k^{(I)} (q\bar q|\varphi\varphi)$ and define it in the following way.
  Using the definition of $\tilde g_1(P)$ (\ref{3.10}) and leaving for $ \tilde g_1$ only the combination $\frac{M^2_n}{M^2_n-E^2}$, one can associate the transition coefficient with the  following combination
\be k^{(I)} (q\bar q|\varphi\varphi)= V_{\bar q\bar q| \varphi\varphi} V_{ \varphi\varphi|\bar q\bar q } = (V_{  q\bar q| \varphi\varphi})^2 = \frac{C^2_i M^2(\lambda) (f_s^{(n)})^2}{f^4_\varphi}, ~~ f_\varphi = f_\pi,f_K.\label{28}\ee
Here the coefficient $C_i$ can be found  from (\ref{20} - \ref{25}).   Introducing notation $C_i = C^I_{\rm meson,meson}$ one obtains from (\ref{5}) and (\ref{20} - \ref{25}),
\be (C^{(0)}_{\pi\pi} )^2 =3; ~~(C^{(0)}_{K\bar K} )^2=2;~~ (C^{(1)}_{K\bar K} )^2=2;~~ \left(C^{(1)}_{\pi\eta} \right)^2=\frac{2}{\sqrt 3}.\label{29}\ee

We start with the one-threshold situation and choose the channel $\pi\pi$, neglecting its connection to $K\bar K$. In this case  one has the following basic elements, with notation $g_2(\pi\pi, E) \equiv g_\pi, ~~ \tilde g_1(E,\veP  =0) = g_1 $, where we keep the lowest pole $M_1$,with the notation $V_{q\bar q|\pi\pi}=V_1{\pi}=V_{\pi  1}$

\be k^{(0)} (n\bar n|\pi\pi) = (V_1{\pi})^2 = \frac{(C_{\pi\pi }^{(0)})^2 M(\lambda) (f^{(1)}_s)^2}{f^4_\pi}=(V_1\pi)^2 ,       ~~ g_1 = \frac{M_1^2}{M^2_1 - E^2},  \label{34a}\ee
and the infinite series for the total $\pi\pi$ Green's function reads, see Fig.3
\be G_{\pi\pi} =g_\pi+ g_\pi V_{\pi 1} g_1 V_{1\pi} g_\pi + g_\pi V_{\pi 1} g_1 V_{1\pi} g_\pi V_{\pi 1} g_1 V_{1\pi} g_\pi+..., \label{35a}\ee
which can be summed up in the form
\be G_{\pi\pi} = g_\pi + g_\pi V_{\pi 1} \frac{1}{1- g_1 V_{1\pi} g_\pi V_{\pi 1}} g_1 V_{1\pi}g_\pi.\label{36a}\ee

For the $\pi\pi$ scattering amplitude $f_\pi(e)$, since $g_\pi$ does not contain $\pi\pi$ interaction, one can define
$$ G_{\pi\pi} = g_\pi + g_\pi f_\pi(E) g_\pi,$$ and one has
\be f_\pi (E) =\frac{1}{16\pi} V_{\pi 1} \frac{1}{1-                                                                                                                                                                                                                                                                                                                                                                                                                                                                                                                                                                                                                                                                                                                                                              \Box_\pi } g_1V_{1\pi},\label{37a}\ee
where we have defined the 4-term  code $\Box_\pi \equiv  g_1 V_{1\pi}g_\pi
 V_{\pi 1}=g_1g_\pi k^{(0)}(n\bar n|\pi\pi)$.

In an analogous way one can define the one-channel $K\bar K$ Green's function and amplitude
 \be f_K (E) = \frac{1}{16\pi}V_{K 1}  \frac{1}{ 1-                                                                                                                                                                                                                                                                                                                                                                                                                                                                                                                                                                                                                                                                                                                                                            \Box_K } g_1V_{1K},\label{38a}\ee
where
\be \Box_K = g_1  g_K   k^{(0)}(n\bar n|K\bar K),   ~~ k^{(0)}(n\bar n|K\bar K) =\left(\frac{C_{K\bar K}^{(0)} M(\lambda) f^{(1)}_s}{f_K} \right)^2.
\label{39a}\ee

Note, that the $q \bar q$ pole at $E^2=M^2_1$ is cancelled in (\ref{37a}), (\ref{38a}); the only visible  singularity is the unitary cut in $g_\pi$ and $g_K$ respectively.

One can check the unitarity of both amplitudes $f_\pi$ and $f_K$,
\be \operatorname{Im} f_\pi (E) = \frac{2k_\pi}{  E} |f_\pi (E)|^2\label{40a}\ee
and the  similar form for $f_K$ is valid with  replacement $\pi\to K$.

One can also find the position of the pole in the amplitude $f_\pi(E)$   from the  denominator in (\ref{37a}),  $\Box_\pi (E) =1$.
One has
\be g_\pi k^{(0)}(n\bar n|\pi\pi)\frac{M^2_1}{M^2_1-E^2}=1, ~~ g_\pi (E) = \operatorname{Re} g_\pi +i \operatorname{Im} g_\pi .\label{41a}\ee
We take here $M_1=1.05$ GeV as follows from (\ref{3.9}).

In the real part of $g_\pi(E)$ the cut-off $N$ is taken at large momenta in (\ref{31}), equal  to the minimal length $\lambda, N=1/\lambda$, which yields $(N=1$ GeV).
\be g_\pi (E)  \approx 0.033 + i 0.02 \sqrt{1-\frac{0.078}{E^2}}.\label{42a}\ee

Inserting in (\ref{41a}) for $I=0$, $f_\pi =93$ MeV, $f_s^{(1)} =125$ MeV (see Appendix 1 for the discussion of $f_s^1$ and Appendix 3 for the pole position in the complex plane), and $M(\lambda=1$ GeV$^{-1}) =180$ MeV, $(C^{(0)}_{\pi\pi})^2 =3$, one obtains the equation
\be E^2=M^2_1 (1-20.3 g_\pi(E)),\label{43a}\ee
or using (\ref{42a}), one obtains the   resonance position $ E_\pi = (0.67-i0.45)$ GeV $(\lambda =1$ GeV$^{-1}$). As a result, varying $\lambda$ in the range $(1\div1.5)$ GeV$^{-1}$ one obtains the    resonance parameters

\be E_\pi = (0.6 \div 0.8)~ {\rm GeV} -i(0.2\div 0.45)~ {\rm GeV}.\label{44a}\ee
 
Note, that the resonance appears on the second sheet of the complex plane with respect to the $\pi\pi$ threshold, as it is explained in Appendix 3.

 This  can be  favorably compared with the experimental values    $f_0 (500), E= (400-550)  $ MeV,  $  \Gamma = 400\div 700$ MeV \cite{1}. Note that  we have obtained these values, however,  with several  simplifying approximations, including the neglect of higher levels in $g_1$, possible coupling with the  $K\bar K$  channel and  notably  neglecting the  $ 4\pi, 6\pi,$... vertices of the  chiral theory, which imply the $\pi\pi$ interaction in $g_\pi (E)$. Therefore, the  resonance position and the width are subject to essential changes, if one takes this interaction into account. In particular, one can notice, that the resonance position (\ref{44a}) is some $(150-200) $ MeV higher, than in experiment.

We now turn to the $K \bar K$ channel, again neglecting connection to the $\pi\pi$ channel and keeping only the lowest mass eigenvalue $M_1= 1.05$ GeV in $g_1(E)$.
 Inserting in (\ref{38a}) $ (C_{K\bar K}^{(0)})^2 =2$, $ f_K =115$ MeV, $M(\lambda=1 $ GeV$^{-1}$)=180 MeV, and $f_s =125$ MeV, one obtains $k^{(0)} (n\bar n\ K\bar K)=5.8$. From (\ref{38a}) one finds  the equation for the  pole position,  $\Box_K=1$, or

\be E^2=M^2_1 (1-5.8~ g_K(E)),\label{44ab}\ee
where $  g_K (E)$ with the upper limit $N=1$ GeV   in (\ref{32a})

\be g_K (E) =  0.018  + i 0.02 \sqrt{\frac{E^2-4m^2_K }{E^2}},\label{45a}\ee
which yields an approximate position of  the pole
\be E_K =(\lambda=1~{\rm GeV}^{-1})=
(0.984-i0.013) ~   {\rm GeV}.\label{46a}\ee
One can see that the pole $E_K$   can be associated  with the standard  $f_0 (980) $ \cite{1}
\be  M(f_0 (980)) = (990 \pm 20) ~{\rm MeV},~ \Gamma = (10\div 100) ~ {\rm MeV},\label{47a}\ee
while the obtained width is inside the allowed region.  It is interesting that  in this case the cut-off $\lambda$ in the range ($0.5  \div 2)$ GeV$^{-1}$ brings about only few percent change in the resulting resonance parameters.
Taking into account the approximations  made
above, this agreement can be  considered as reasonable, however, one should take into account, that both channels $\pi\pi$ and $K\bar K$ should be connected, as it is seen in the experimental measurements of the  ratio for $f_0(980),   \frac{\Gamma (K\bar K)}{\Gamma(\pi\pi)}  = 0.69 \pm 0.32$ \cite{1}.

The standard way to include  the $\varphi\varphi $ channel coupling is to write for the amplitudes $\hat f_{\alpha \beta} = \left( \begin{array}{ll} f_{\pi\pi} & f_{\pi K}\\f_{K\pi} & f_{KK} \end{array}\right)$ the $K$ matrix form,
\be \hat f^{-1}=  16\pi\left( \begin{array}{ll} \frac{1-\Box_\pi}{w_\pi}  &a\\b&  \frac{1-\Box_K}{w_K} \end{array}\right),~~
{w_\pi}= V_{\pi 1} g_1 V_{1\pi}, ~ w_K=V_{k_1g_1} V_{iK}.\label{49b}\ee

As a result  one obtains
 \be  \hat f = \frac{1}{16\pi}\frac{ \left( \begin{array}{ll} \frac{1-\Box_K}{w_\pi}  &-b\\-a&  \frac{1-\Box_\pi}{w_K} \end{array}\right)}{    \frac{1-\Box_\pi}{w_\pi} \cdot    \frac{1-\Box_K}{w_K}-ab},\label{49a}\ee
and in the limit $ab=0$ one returns to the  two  independent channels.

One can   check  that the amplitudes $f_{\alpha \beta},  \alpha, \beta =\pi \pi, K \bar K$  satisfy the unitarity relations with the  normalization  factor  $\operatorname{Im} g_{\pi, K} = \frac{k_{\pi, K}(E)}{8\pi  E}$. In particular  for  the $f_{K\bar K}$ one  has  in this channel coupling  ($CC$) form

 \be 16\pi f_{K\bar K} = \frac{(1-\Box_\pi) w_K}{(1-\Box_\pi) (1-\Box_K)-ab  w_K w_\pi}\label{50a}\ee
 Estimating the $w_K, w_\pi$ one finds that the CC can   affect the positions and the widths  of the uncoupled resonances (\ref{44a}), (\ref{46a}) and therefore this point should be studied in more detail.

 We shall start with  the $f_{\pi\pi}$ amplitude, which can be written   as follows
 \be 16\pi f_{\pi\pi} = \frac{\frac{1}{w_K}-g_K}{\left(\frac{1}{w_\pi} - g_\pi\right) \left(\frac{1}{w_K} - g_K\right)-ab},\label{51a}\ee
  and we can  rewrite (\ref{51a}) as follows using $\gamma = 40.4~ ab$ and the properly normalized amplitudes $\operatorname{Im}  f_\pi^{(0)} = \frac{2k}{E} |f_\pi^{(0)}|^2$

  \be f_\pi^{(0)} =\frac{1}{\frac{16\pi}{k} \frac{(E^2_\pi-E^2)}{M^2_1}- \frac{\gamma M^2_1}{E^2_K - E^2}}\label{57a}\ee
  with \be E^2_\pi= M^2_1 \left(1-k^{(0}(n\bar n|\pi\pi)g_\pi \right), ~ E^2_K = M^2_1 \left(1-k^{(0}(n\bar n|K\bar K)g_K\right).\label{58a}\ee
Analogously for $f^{(0)}_K$ one has
\be f^{(0)}_K= \frac{1}{\frac{16\pi}{k M^2_1} (E^2_K-E^2) - \frac{2.34 \gamma M^2_1}{E^2_\pi - E^2}} .\label{58}\ee


One  can estimate the ratio of imaginary parts of the first and the second term in the  denominator of (\ref{58}), which yields the order of magnitude of the ratio of $\Gamma_{K\bar K} (f(980))$ and $\Gamma_{\pi\pi} (f(980))$ at $E=1.00$ GeV,
\be \frac{\Gamma_{\pi\pi} (f(980)) }{\Gamma_{K\bar K} (f(980))}\cong \frac{ 2.4 \gamma  \sqrt{ E^2 -4m^2_\pi}}{\sqrt{E^2-4m^2_K}}\approx 17 \gamma\label{59}\ee
and one can see that this ratio is around $1$ for $\gamma =0.05$, found in the next section by comparison with data. The resulting pole is near the $K\bar K$ threshold and satisfies the criteria of the $f^{(0)}(980)$ resonance.

\section{Results and discussion}

From (\ref{57a})  one can see that the amplitude $f^{(0)}_\pi$ can be expressed via the Green's functions $g_\pi$ and $g_K$  with the only parameter $\gamma$,  responsible for the coupling of channels $\pi\pi$ and $K\bar K$.
Note, that the only parameter $\lambda$ enters both $q\bar{q}-mm$ coupling $k^{(0)}$ and the real parts of both  Green's  functions , which start and finish at the same distance $\lambda$ and therefore contain the legitimate factor  $N\sim 1/\lambda \sim O(1\text{ GeV})$. In the present paper we have chosen $\lambda$ in the narrow interval around $1$ GeV$^{-1}$, which has yielded reasonable results. In the subsequent paper \cite{X} it was shown, that $\lambda$ can be defined from the stationary point of the transition coefficient $k^{(0)}$ and indeed has value near $0.2$ Fm $=$ $1$ GeV$^{-1}$.

The main  result  of our approach, based on the CCL (\ref{1}), is that the $q\bar q$ pole at $1$ GeV can provide only one resonance, when connected with one threshold, and we need $\pi\pi-K\bar{K}$ channel coupling to produce two quark-chiral resonances: $f_0(500)$ due to coupling $n\bar{n}-\pi\pi$, and $f_0 (980)$ due to coupling $n\bar{n}- K\bar{K}~(n=u,d)$. This is a  feature of our quark-chiral Lagrangian, and is obtained from the infinite sums of products of $\Box_\pi$ and $\Box_K$. Starting with uncoupled $\pi$ and $K$ channels, it is interesting that the $\pi\pi$ pole, produced by the $q\bar{q}$ pole obtains without the $\pi\pi$ interaction, which is governed by the chiral Lagrangian, however, far from experimental position, and then one needs direct (not via $q\bar{q}$) $\pi\pi$ interaction to bring resonance to the realistic values, which can be obtained directly by the analysis in \cite{19**, 39, 40}. The possible reason is that the low energy physics is only mildly connected to the higher  $f^{(0)}$ resonance physics,  but strongly affects low and intermediate energy region, including  $f_0(500)$ position, and  we have neglected at the first stage the low energy $\pi\pi$ interaction given by the chiral Lagrangian.

Therefore, in our general two-channel form we include a possible modification of real   and imaginary part of $g_{\pi}$ due to direct $\pi\pi$ or $K\bar{K}$ interaction, which is contained in the term $E^2_{\pi}$ in (\ref{58a}), which leads to the following two-channel form, generalizing (\ref{57a}),

\be f_\pi  = \frac{1}{2.67 \frac{\tilde E^2_\pi - E^2}{M_1^2 }} - \frac{\gamma M^2_1}{E^2_K -E^2}; ~~ \tilde E^2_\pi  = M^2_1 \left(1-x(E) -iy(E)\sqrt{\frac{E^2-4m^2_\pi}{E^2}}\right) \label{5.4}\ee

\be F_2(E) = 0.96-0.043 ~i\sqrt{1-\frac{0.975}{E^2}} \theta (E^2-0.975) - E^2\label{5.5}\ee
Here  $x(E) = k  \operatorname{Re}  g_\pi (E), ~ y= k \operatorname{Im} g_\pi (E)$ are found by fitting the  resulting  curves of  $ \operatorname{Re} f^{(0)}_\pi,  \operatorname{Im} f^{(0)}_\pi$ to the data of \cite{39,40}.
 Parameters of $x(E), y(E)$ are given in the Appendix A3.

The two curves $f_\pi(E)$,  obtained by the fitting of $g_\pi(E)$, are shown in Figs. 4 and 5 by   the solid lines together  with the curves from the paper of  Pelaez  et al.  \cite{39}, obtained in the  course of the analysis in \cite{40}. In the same figures we show the dashed line curves obtained from (\ref{57a}),(\ref{58a}) with the free $\pi\pi$ Green's function. As one can see, our $\operatorname{Re} f_\pi^{(0)}(E)$ and $\operatorname{Im} f_\pi^{(0)}(E)$  for the  free case are in a qualitative agreement with the results of \cite{39}, with an  exclusion of the region of relatively small energies, $E < 0.5$ GeV.

This  means that the  $\pi\pi $  interaction is  important in this  region and the  approximation of the free $\pi\pi$ Green's  function should be modified by inclusion of the purely chiral  interactions at least for the lowest  $f_0(500)$ resonance.

This is  well illustrated by the calculation of the position of the resonance $f_0(500)$, which was obtained above in the free $g_\pi$ case at $(0.67-i0.45)$ GeV, while from \cite{39} ( the dotted lines in Figs 5,6) the  resonance position is at $(0.457-i 0.259)$ GeV.
\begin{figure}
\begin{center}
  \includegraphics[width=8cm, ]
  {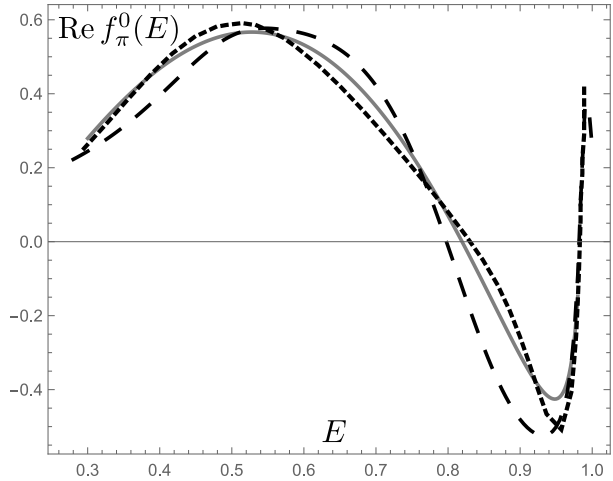}

 \end{center}
  \caption{$\operatorname{Re}  f^{(0)}_\pi(E)$ as a function of $E$ in GeV from Eq. (\ref{57a}) (grey bands) in comparison  with the resulting curves from the Pelaez et al.   \cite{39,40} (broken lines)  comprising the $\pi\pi$ data. }

\end{figure}


\begin{center}

\begin{figure}
\begin{center}
  \includegraphics[width=8cm, ]
  {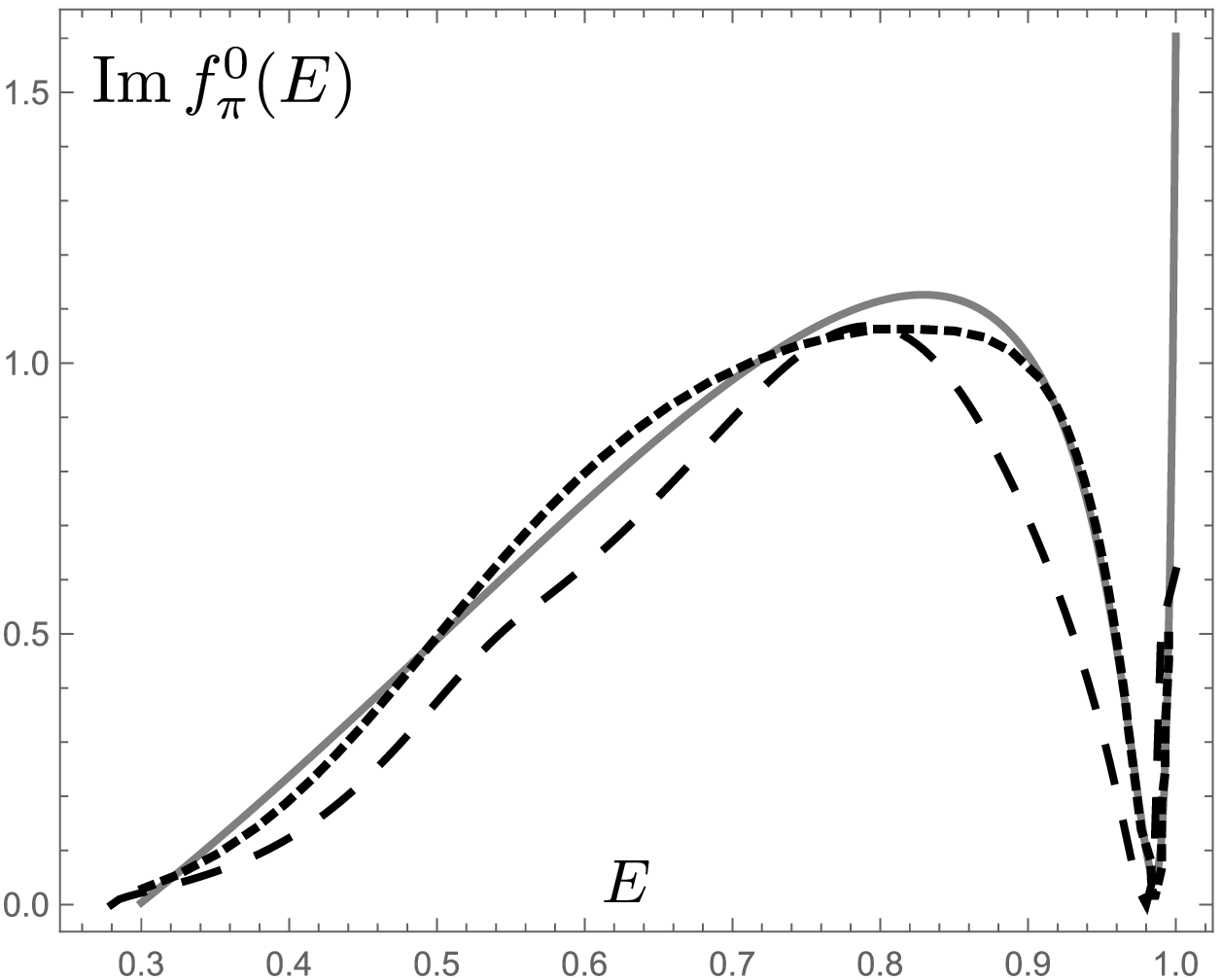}

\end{center}
  \caption{The same as in Fig. 4, but for $\operatorname{Im} f_\pi^{(0)}  (E)$. }

\end{figure}
 \end{center}

However, it was not the purpose of our study to reproduce exactly the $\pi\pi$ interaction amplitude in the  whole region (280, 1000)  MeV, but  rather to   discover the dynamical mechanism, producing the lowest scalar-isoscalar mesons $f_0(500)$ and  $f_0(980)$. As shown, at the first stage  this mechanism  can be reduced in its basic part  to the interaction of the $q\bar q$ and  free meson-meson channels, given by our  quark-chiral interaction in the CCL, Eq. (\ref{1}). Indeed,  this interaction  provides the reasonable coupling $V_{q\bar q|\pi\pi}$ and $V_{ q\bar q|K\bar K}$, in addition to the values of the $q\bar q$ Green's functions and the  corresponding poles $M_n (q\bar q)$.  In our case the lowest pole $M_1(q\bar q)$ at 1 GeV produces a wide resonance $f_0(500)$ in ``collaboration'' with the $\pi\pi$ Green's function and  the  $\pi\pi$ threshold, and a more narrow resonance $f_0(980)$ in ``collaboration'' with the $K \bar K$  Green's function and the threshold. The interaction of these two channels, strongly shifted  in energy from each other, which is outside our simple $q\bar q$-meson-meson model, only slightly modifies their individual properties, as can be seen comparing one-channel and coupled-channel characteristics.

 As the second stage one should take into account  the chiral interactions $(\pi\pi, K\bar K)$  to obtain  the relativistic  $\pi\pi$ and $K\bar K$ amplitudes. This stage is   especially essential for the  determination of the $f_0(500)$ pole parameters, as can be seen  from the  analysis in \cite{39}, $E=(457\pm 10)$ MeV$-i 279 $ MeV,  which agrees with the results of \cite{19**}. This result   disagrees with our estimate (\ref{44a}), where the $\pi\pi$ interaction was disregarded. At the  same time the characteristics of the $f_0(980)$ in \cite{39} and in our case, Eq. (\ref{46a}) are similar.  This leads to the  conclusion, that the  accurate determination of lowest   resonances, much below 1 GeV, requires the proper account of the $\varphi\varphi$ interaction, which can be done combining the    formalism of  \cite{19**, 39}, with our $q\bar q-\varphi \varphi$ approach.

This is the main  concrete result of this paper, however, the general mechanism, described above, leads to many further possible discoveries.

At this point one can immediately ask: if the same $q\bar q$ level can create several resonances, accounting for the coupling between $\varphi\varphi$ channels, what happens with $a_0(980)$ resonance, which can decay both to $\pi\eta$ and $K\bar K$, but in experiment one can see only one broad resonance near the $K\bar K$ threshold. Now we apply our technique to this case to understand the difference between the situation with $a_0(980)$ on one hand and $f_0(500), f_0(980)$ on another.

To this end we shall try to find separate resonances in the $\pi\eta$ and $K\bar K$ channels and  write, as in (\ref{43a}), the resulting equation for the position of the assumed resonances $E^{(1)}(\pi\eta )$ and $E^{(1)} (K\bar K)$, where the upper index refers to the isospin $I=1$.

\be (E^{(1)} (\nu))^2 = M^2_1 (1- k^{(1)}(n\bar n|\nu) g_\nu (E));~\nu=\pi\eta, K\bar K. \label{63}\ee
Now using (\ref{34a}) and  (\ref{21})-(\ref{25}) one can write:
\be k^{(1)}(n\bar n|\pi\eta)\approx  V_{\pi\eta,1} V_{1,\pi\eta} = (C^{(1)}_{\pi\eta})^2\frac{M^2(\lambda) (f_s^{(1)})^2}{f^2_\pi f^2_\eta} =   4.69\label{64}\ee
\be k^{(1)}(n\bar n|K\bar K) = \frac{(C_{K\bar K}^{(1)})^2 M^2(\lambda) (f_s^{(1)})^2}{f^4_K}=2.9\label{65}\ee
(We  have neglected the difference between $f_\pi$ and $f_\eta$ for a rough estimate, in  the real case the coefficient in (\ref{64}) is smaller).

Now $g_{\pi\eta} (E)$ has a smaller real and imaginary parts, (see Appendix 2 for details) as compared with $g_{\pi\pi}(E)$ Eq. (\ref{42a}), while $g_{K\bar K}(E)$ is the same as was used before, see Eq.(\ref{45a}). As a result, the solution of Eq. (\ref{63}) gives two resonances
\be E^{(1)}(\nu) = M_1 (1- \bar a_\nu - i \bar b_\nu),\label{66}\ee
where a rough estimate yields
\be \bar a_{\pi\eta} \cong0.05, ~~ \bar b_{\pi\eta} \approx 0.05\label{67}\ee
\be \bar a_{K\bar K} = 0.022, ~~ \bar b_{K\bar K} =0.04 \sqrt{\frac{E^2-4m^2_K}{E^2}}.\label{68}\ee
One should take into account that $M_1(I=1)\approx M_1(I=0) = 1.00~{\rm GeV}$ and obtains $E^{(1)}(\pi\eta) \cong (1.05) $ GeV, while
$E^{(1)} (K\bar K) \cong \left(1.04 - i 0.02 \sqrt{\frac{E^2-4m^2_k}{E^2}}\right)$ GeV, and $E^{(1)} (\pi\eta)$ is  on the second sheet with the $\pi\eta$ threshold, while $E^{(1)}(K\bar K$) on the second sheet with the  $K \bar K$ threshold.

One should  however take into account the $\pi\eta-K\bar K$ channel coupling, which can rearrange the position of the poles, as it was found recently in the lattice calculations \cite{12'}.

Thus, one can see that the  displacements of both resonances are small,  being of the order of the width of resonances. This might be  the reason why in experiment one actually observes one resonance  $a_0(980)$ near $1$ GeV with two decay modes, while in the $I=0$ channel with larger couplings $k^{(0)}  (n\bar n|{\pi\pi }) $ and more distant $\pi\pi$ and $K\bar K$ thresholds one observes two distinct resonances, and this example gives an additional support for our theory.


\section{Conclusions and an outlook}

First of all, comparing our approach with other models, one should stress that we neglect any direct $\varphi\varphi$
interaction in the first step, described in the paper. Therefore, all details of this interaction, as well as $q\bar{q}-q\bar{q}$  interaction, in particular crossing symmetry, the left-cut singularities etc are missing in this first step. As a second, and a more complicated step, one should take into account all details of the $\varphi\varphi$ interaction, e.g. as in the dispersive methods or in unitarized chiral model interaction.

Summarizing, the method suggested above as a first stage, has a general character and can be applied to any systems, consisting of several components, which can transform one into another. The only information, needed to describe the  properties of such mixed systems, is  the  spectral properties of each component and  transition coefficients. In the case of the charmonium system this method has given a first explanation of the resonance $X(3872)$ \cite {28'''}. In the case of the quark-chiral system, $q\bar q-\varphi\varphi$, this method uses the information  given by the FCM approach plus quark-chiral CCL Lagrangian (\ref{1}). As it is, our method suggests a possible solution of the old-standing problem of $f_0(500)$, $f_0(980)$ and $a_0(980)$
associating these resonances with $(n=1)$ $q\bar q$ $^3P_0$ states. 

As applied to the lowest scalar resonances, we have shown that  the resonances $f_0(500)$ and $f_0(980)$, as well as the $a_0(980)$ resonance, can be connected with the use of CCL, to the  $n=1,\,M\approx 1$ GeV  $q\bar q$ resonance calculating explicitly the transition coefficients and consequently the partial widths.  Then several questions arise:
\begin{enumerate}
\item Since we have connected $f_0(500)$, $f_0(980)$  with  one $q\bar q$ state -- the $^3P_0$ ground state $n\bar n$ with mass around $1$ GeV, one should consider the next $q\bar q$ state, $M_2(1474)$ as an excited $q\bar q$ state with $n=2$, in contrast to an  accepted view (see \cite{1}) that this latter is a ground state. It is interesting to study consequences of this assignment.
\item What will be the  result for excited $q\bar q-\varphi\varphi$ states, e.g. with $M_2= 1474$ MeV,   in connection with the same $\varphi\varphi$ thresholds and can one expect more additional resonances below $M_2$ ?
\item It is clear that  taking into account the  full sum $\sum_n \frac{M^2_n}{M^2_n -E^2}$ one meets with divergences and with the necessity of renormalization. This probably  can be  treated  in the  spirit of the formalism, developed in the method of Matrix Product States (MPS), see \cite{41} for reviews.

\item  We have considered above only one $q\bar q$ channel. However, for the  $K\bar K$ system  the $s\bar s$ channel provides bound states starting with $M_1\cong 1400 $ MeV, just near the first excited $n\bar n$ state. Therefore, for the $K\bar K$ system one should take into account both $n\bar n$ and $s\bar s$ states, which  requires an extension of our  method  with inclusion of  several $q\bar q$ and one or more $\varphi\varphi$ channels to  explain several extra resonances in the region 1300-1700 MeV, observed in experiment \cite{1}.
\end{enumerate}

These topics have been recently studied in \cite{X} as our subsequent paper and it was shown, how excited $q\bar q$ states produce higher scalar resonances using the explicit method, discussed in the present paper. In addition, also $n\bar s$ and $s\bar s$ systems have been considered in connection with the corresponding $mm$ thresholds and the same picture of the pole shifts as in the present paper was found. From this point of view our approach helps to clarify the old problem of scalar resonances  both in the ground and first
 excited states, leaving the question: what parts in lowest scalars are occupied by $q\bar q$ and direct $\varphi\varphi$ interactions for a future investigation. \bigskip

This work was done in the  frame of the scientific project, supported by the Russian Science Foundation
grant number 16-12-10414.  The authors are indebted to A.M. Badalian for many discussions, suggestions and details, which are used in the paper. Discussions with Yu.S. Kalashnikova and Z.V.Khaidukov are gratefully acknowledged.





\section*{Appendix A1. Decay constants of scalar mesons}

\setcounter{equation}{0} \def\theequation{A1.\arabic{equation}}

 In the framework of the path-integral formalism the decay constants of the
 $q\bar q$ meson states can be defined as in  \cite{14,28}

\be (f_\Gamma^{(n)})^2=\frac{2N_c \lan Y_\Gamma\ran | \varphi_n (0)|^2}{
\omega_1 \omega_2  M_n}, \label{A1}\ee where  $\omega_1, \omega_2$ are average
energies of quarks with masses $m_1$ and $m_2 , M_n$  is the mass of the meson,
$\varphi_n (r)$ is the (relativistic) meson wave function of the relative
distance $r$, while $\lan Y_\Gamma\ran$ is
 \be
4Y_\Gamma = \operatorname{tr} ((m_1-\hat D_1) \Gamma(m_2-\hat D_2)\Gamma )=  \operatorname{tr} ((m_1-i\hat
p_1) \Gamma(m_2+i \hat p_2)\Gamma).\label{A2}\ee

Here $\Gamma$ is the vertex operator, for the scalar particle $\Gamma_s=1$, but
the momentum operators $\hat p_i$ are acting on the wave function $\varphi_n
(r)$, namely $ip_i\varphi_n (r) = \partial_i \varphi_n (r).$ In our case \be
\lan Y_s\ran |\varphi_n (0)|^2 = (m_1m_2 -\omega_1\omega_2 -
\hat{\vep}\hat{\vep}') |\Psi_S (0)|^2\to (\partial_i \Psi_S (\ver) \partial'_i
\Psi^*_S (\ver'))_{r\to 0, r'\to0}.\label{A3}\ee Since $\Psi_S(\ver)$ is
\be\Psi_S(\ver)  = \sum \chi_{1M_1} \tilde Y_{1m_2} \frac{\varphi(r)}{r}
C^{00}_{1m_1, 1m_2}\label{A4}\ee and $\tilde Y_{1m}\equiv rY_{1m}$, after
summation over spin projections one finds \be \partial_i \Psi_S (\ver )
\partial'_i \Psi^*_S(\ver')= \partial_i \partial'_i \frac{1}{4\pi} (xx' +
yy'+zz') = \frac{1}{4\pi} \label{A5}\ee where we have taken into account, that
the subscript $i$  refers to a fixed momentum direction. As a result one
obtains

\be (f^{(n)}_S)^2 = \frac{2N_c (R'_{nP} (0))^2}{ 4\pi \omega_1
\omega_2M_n},\label{A6}\ee where $R'_{nP} (0) =
\left(\frac{\varphi_n(r)}{r}\right)_{r\to 0}$. Estimated in  the same way as in \cite{14,28}
for the $1P$ scalar state one has $R'_{1P}(0)=0.086$ GeV$^{5/2},
\omega_1=\omega_2= 0.448$ GeV  \cite{42} and according to (\ref{A6}) one obtains \be
(f_S^{(1)})^2 =0.01568~{\rm GeV}^2, ~~ f_S^{(1)} =0.125~{\rm GeV}.\label{A7}\ee

For the first excited state, $2P$, one has for the scalar state $\omega(2P)
\cong 0.5$ GeV, $ R'_{2P} (0) =0.0817$ GeV$^{5/2}$, $M(2P) =1.474$ GeV \cite{42}.

As a result one obtains from (\ref{A6})

\be (f_s^{(2)})^2 =0.00865~{\rm GeV}^2, ~~ f_s^{(2)} =0.093~{\rm
GeV}.\label{A8}\ee

\section*{Appendix A2. Meson-meson Green's functions}

\setcounter{equation}{0} \def\theequation{A2.\arabic{equation}}

The relativistic Green's function of two scalar mesons with the total momentum $P$ can be written  in the  Euclidean space-time as
\be g(P) = \int \frac{d^4p}{(2\pi)^4} \frac{1}{[(P-p)^2+m^2_1](p^2+m^2_2)}.\label{A3.1}\ee

Integrating over $dp_4$ in the c.m. frame,
 $\veP =0$, one obtains, with $P_4=iE$, and $m_1>m_2$,
$$
\operatorname{Re} g_{12} (E) = \int^N_0 \frac{p^2dp}{4\pi^2}\times $$ \be\times\frac{E(\sqrt{p^2+m^2_1}+\sqrt{p^2+m^2_2})+ m^2_1-m^2_2}
{\sqrt{p^2+m^2_1} \sqrt{p^2+m^2_2} [(\sqrt{p^2+m^2_1}+\sqrt{p^2+m^2_2})^2-E^2]
(E+\sqrt
{p^2+m^2_1}-\sqrt{p^2+m^2_2})} \label{A3.2}
\ee
Here we have introduced the cut-off $N$ in momentum $p$.
\be \operatorname{Im} g (E) = \frac{1}{16\pi} \frac{\sqrt{(E^2-(m_1+m_2)^2)(E^2-(m_1-m_2)^2)}}{E}.\label{A3.3}\ee
In the equal mass limit one obtains
\be \operatorname{Re} g (E) = \int^N_0 \frac{p^2 dp}{8\pi^2 \sqrt{p^2+m^2} (p^2+m^2-E^2/4)},\label{A3.4}\ee
which  for $E^2=4m^2$ reduces to a simple answer
\be \operatorname{Re} g (2m) = \frac{1}{8\pi^2}\int^N_0 \frac{dp}{\sqrt{p^2+m^2}}=\frac{1}{8\pi^2} \ln \frac{N+\sqrt{N^2+m^2}}{m}.\label{A3.5}\ee
For $E^2=4m^2-4\Delta, ~~ \Delta>0$ one has instead of (\ref{A3.5})
\be \operatorname{Re} g (E) = \frac{1}{8\pi^2}\int^N_0 \frac{dp}{\sqrt{p^2+m^2}}-\frac{\Delta}{8\pi^2} \int^N_0  \frac{dp}{\sqrt{p^2+m^2}(p^2+\Delta)}.\label{A3.6}\ee
and for $m^2\gg\Delta$ the last integral in (\ref{A3.6}) can be written as
\be \Delta \operatorname{Re} g (E) \cong - \frac{\sqrt{\Delta}}{16\pi m} \theta \left(m^2- \frac{E^2}{4}\right).\label{A3.7}\ee
Note, that $\Delta \operatorname{Re} g (E)$ is much smaller than $\operatorname{Re} g (2m)$, Eq. (\ref{A3.5}) and can be neglected in the first approximation.

\section*{Appendix A3. Position of new poles in the complex plane}

\setcounter{equation}{0} \def\theequation{A3.\arabic{equation}}

One can write the equation (\ref{43a}) for the $\pi\pi$ pole as
\be E^2=M_1^2\left( 1-\operatorname{const} (\operatorname{Re}g_\pi(E)) + i \operatorname{const}(\operatorname{Im}g_\pi(E)) \right) \label{A3.1} \ee

Writing $\operatorname{Im} g_\pi(E) = \operatorname{const} p(E)/E$, where $p(E)=\sqrt{E^2-4m_\pi^2}$, one can rewrite (\ref{A3.1}) as
$E^2 = E_0^2 - \frac{i b p(E)}{E}$, or else expressing $E^2$ via $p^2$ one has
\be p^2(E) + \frac{i b p(E)}{\sqrt{p^2(E)+4m_\pi^2} - p_0^2} = 0 \label{A3.2} \ee
 where $b$ and $p_0$ are constants. Starting with small $b$ one will have approximately $p(E)=p_0$,and in the next order having $\sqrt{p^2(E) + 4m_\pi^2}=\sqrt{P_0^2 + 4m_\pi^2}$, one solves the quadratic equation for $p(E)$ as 
\be p(E)=-\frac{ib}{2\sqrt{p^2(E)+4m_\pi^2}} \pm \sqrt{\frac{-b^2}{4(\sqrt{p^2(E)+4m_\pi^2})^2} +p_0^2} \label{A3.3} \ee
 One can do next orders of approximations following the motion of the root, starting with the position (\ref{A3.3}).
 Another way is the direct solution of the equation (\ref{A3.2}) which is cubic in $p^2(E)$ choosing the correct root to be consistent with (\ref{A3.3}).As it is seen in (\ref{A3.3}) the sign of imaginary part of $p(E)$ is negative, implying that the pole is on the second sheet in the $E$-plane, corresponding to the $\pi\pi$ threshold.

\end{document}